\documentclass[prl,showpacs,preprintnumbers,amsmath,amssymb,twocolumn,superscriptaddress,longbibliography]{revtex4-1}
\usepackage{graphicx}
\def\be{\begin{equation}}
\def\ee{\end{equation}}
\def\bea{\begin{eqnarray}}
\def\eea{\end{eqnarray}}

\begin{document}

\title{Evidence of Nodal Line in the Superconducting Gap Symmetry of Noncentrosymmetric ThCoC$_{2}$} 

\author{A. Bhattacharyya}
\email{amitava.bhattacharyya@rkmvu.ac.in} 
\address{Department of Physics, Ramakrishna Mission Vivekananda Educational and Research Institute, Belur Math, Howrah 711202, West Bengal, India}
\affiliation{ISIS Facility, Rutherford Appleton Laboratory, Chilton, Didcot Oxon, OX11 0QX, United Kingdom} 
\affiliation{Highly Correlated Matter Research Group, Physics Department, University of Johannesburg, PO Box 524, Auckland Park 2006, South Africa}
\author{D. T.  Adroja} 
\email{devashibhai.adroja@stfc.ac.uk}
\affiliation{ISIS Facility, Rutherford Appleton Laboratory, Chilton, Didcot Oxon, OX11 0QX, United Kingdom} 
\affiliation{Highly Correlated Matter Research Group, Physics Department, University of Johannesburg, PO Box 524, Auckland Park 2006, South Africa}
\author{K. Panda}
\address{Department of Physics, Ramakrishna Mission Vivekananda Educational and Research Institute, Belur Math, Howrah 711202, West Bengal, India}
\author{Surabhi Saha} 
\affiliation{Department of Physics Indian Institute of Science Bangalore 560012 India}
\author{Tanmoy Das} 
\email{tnmydas@gmail.com}
\affiliation{Department of Physics Indian Institute of Science Bangalore 560012 India}
\author{A. J. S. Machado} 
\affiliation{Escola de Engenharia de Lorena, Universidade de Sao Paulo, P.O.Box 116, Lorena, Sao Paulo, 12602-810, Brazil.}
\author{T. W. Grant} 
\affiliation{Escola de Engenharia de Lorena, Universidade de Sao Paulo, P.O.Box 116, Lorena, Sao Paulo, 12602-810, Brazil.}
\affiliation{Department of Physics and Astronomy, University of California-Irvine, Irvine, CA 92697, USA}
\author{Z. Fisk} 
\affiliation{Department of Physics and Astronomy, University of California-Irvine, Irvine, CA 92697, USA}
\author{A. D. Hillier} 
\affiliation{ISIS Facility, Rutherford Appleton Laboratory, Chilton, Didcot Oxon, OX11 0QX, United Kingdom}
\author{P. Manfrinetti} 
\affiliation{Department of Chemistry, University of Genova, 16146 Genova, Italy}

\date{\today}

\begin{abstract}
The newly discovered noncentrosymmetric superconductor ThCoC$_{2}$ exhibits numerous unconventional behavior in the field dependent heat capacity data. Here we present the first measurement of the gap symmetry of ThCoC$_{2}$ by muon spin rotation/relaxation $(\mu$SR) measurements. Temperature dependence of the magnetic penetration depth measured using the transverse field $\mu$SR measurement reveal the evidence of nodal pairing symmetry. To understand these findings, we carry out the calculations of superconducting pairing eigenvalue and eigenfunction symmetry due to the spin-fluctuation mechanism, by directly implemented the {\it ab-initio} band structures. We find that the system possesses a single Fermi surface with considerable three-dimensionality, and hence a strong nesting along the $k_z$-direction. Such a nesting promotes a superconducting pairing with a $\cos{k_z}$-like symmetry with a prominent nodal line on the $k_z=\pm\pi/2$ plane. The result agrees well with the experimental data.

\end{abstract}


\maketitle


In recent years, noncentrosymmetric (NCS) superconductors have attracted considerable attention in condensed matter physics, both theoretically and experimentally ~\cite {Simdman2017, Bauerbook}. These NCS compounds lack inversion symmetry, which generates an asymmetric electrical field gradient in the crystal lattice and, thereby, produces a Rashba-type antisymmetric spin-orbit coupling (ASOC, energy scale $\sim$ 10-100 meV)~\cite{Rashba,Frigeri,Samokhin,Fujimoto}.  In presence of  ASOC, the spin of a Cooper pair is not a good quantum number, the spin degeneracy of the conduction band is lifted and hence allows for the admixture of spin-singlet and spin-triplet states in Cooper pairs formation~\cite{Frigeri, Samokhin}. If the triplet part is large~\cite{Frigeri1}, the superconducting gap reveal line or point nodes in the order parameter. The discovery of unconventional superconductivity (USc) in CePt$_3$Si with $T_c$ = 0.75 K, has spurred theoretical and experimental works to find out the effect of the lack of inversion symmetry on NSC~\cite{Bauer2,Yogi,Bonalde1}. USc in pressure was observed in NSC heavy fermion (HF)  compounds~\cite{Kimura2,Kimura3,Okuda1}, such as CeTSi$_3$(T = Rh, Ir)~\cite{Sugitani,Kawai1,Wangw} and  CeTGe$_3$ (T = Co, Rh and Ir)~\cite{Honda}. To date, USc properties, e.g., a nodal gap structure and/or a large upper critical field, have been presented in a few weakly correlated NCS superconductors, including Li$_2$(Pd$_{1-x}$Pt$_x$)$_3$B~\cite{Yuan1,Yuan}, Y$_2$C$_3$~\cite{Chen}, Mo$_3$Al$_2$C~\cite{Bauer1},  Mg$_{10}$Ir$_{19}$B$_{16}$~\cite{Bonalde2}  and with special attention for, LaNiC$_2$~\cite{Hillier}.

\begin{figure*}[t]
\centering
\includegraphics[width=\linewidth,height=5cm]{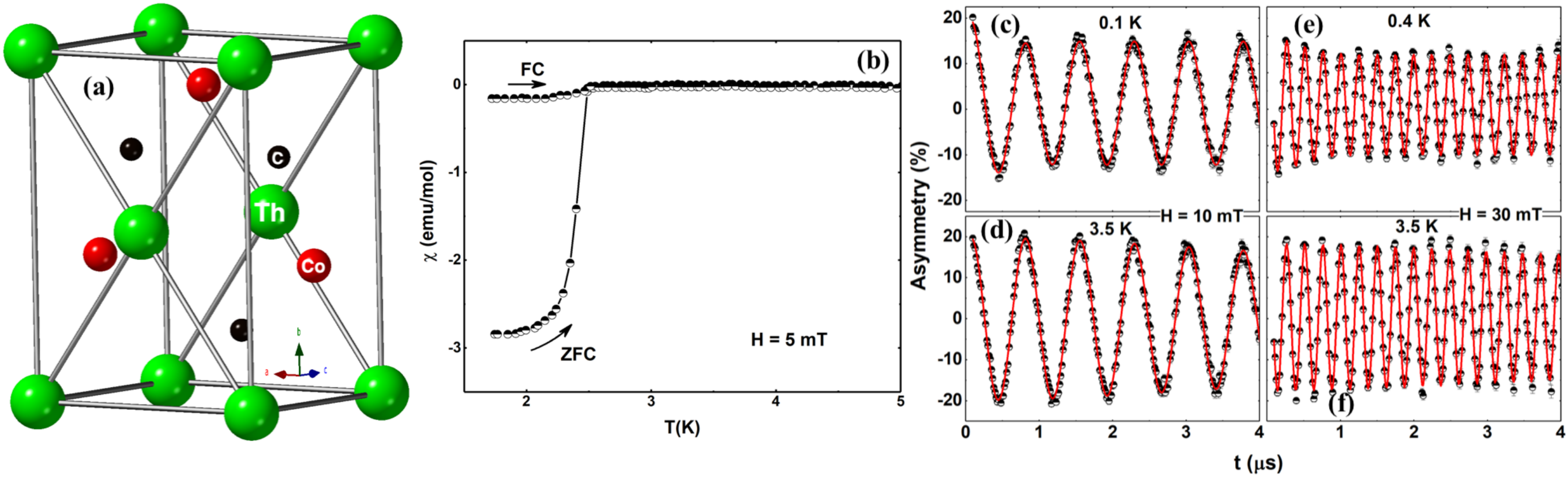}
\caption{(Color online) (a) shows the orthorhombic crystal structure of ThCoC$_{2}$ where the Th atoms (big) are green, the Co  atoms (medium) are red  and the C atoms (small) are black. (b) Temperature dependence of magnetic susceptibility $\chi(T)$ of ThCoC$_{2}$ in ZFC and FC mode.  Transverse field $\mu$SR asymmetry spectra at (c) $T$ = 0.1 K, (d) $T$ = 3.5 K for $H$ = 10 mT and (e) $T$ = 0.4 K, (f) $T$ = 3.5 K for $H$ = 30 mT. The solid lines are fits using  an oscillatory function with a Gaussian relaxation plus a nondecaying oscillation that originates from muons stopping in the silver sample holder. The effect of the flux line lattice can be seen in the top panel as the strong Gaussian decay envelope of the oscillatory function. Above $T_c$, the depolarization is reduced and is due to the randomly oriented array of nuclear magnetic moments.}
\label{fig1}
\end{figure*}

ThCoC$_2$ with $T_c$ = 2.3 K, crystallizes in the NCS CeNiC$_{2}-$type orthorhombic structure with space group  $Amm2$ (No. 38), in which a mirror plane is missing along the $c-$axis~\cite{Bodak,Grant,Grant1}, as displayed in Fig.~\ref{fig1}(a). The upper critical field ($H_{c2}$) vs temperature phase diagram confirms a positive curvature~\cite{Grant}, and the electronic specific heat coefficient at low temperature exhibits a square root field dependence $\gamma(H)\propto \sqrt{H}$, which indicates presence of nodes in the superconducting order parameter~\cite{Chen1,Wright,Yang2,van1} and a potential USc state in ThCoC$_2$. $\gamma(H)\propto \sqrt{H}$ property, was admitted a general sign on $d-$band  and HF superconductors~\cite{Wright, Yang2,van1}. Moreover, examples of positive curvature in $H_{c2}(T)$ phase diagram includes  LnNi$_2$B$_2$C (Ln=Lu and Y)~\cite{Takagi,Shulga}, MgB$_2$~\cite{Takano}, the NCS Li$_2$(Pd,Pt)$_3$B~\cite{Peets}, and the NCS HF CeRhSi$_3$~\cite{Kimura}. The isostructural compound  LaNiC$_2$ displays USc below $T_c$ = 2.7 K with broken time-reversal symmetry (TRS) in $\mu$SR measurements, a nodal energy gap from very-low-temperature magnetic penetration depth measurements, and a suggested multigap superconductivity due to the modest value of the ASOC in LaNiC$_2$~\cite{Hillier,JorgeQuintanilla}. Motivated by these results, in this Letter, we have investigated the compound ThCoC$_2$ using transverse (TF) and zero field (ZF) $-\mu$SR measurements and probed its superconducting state, which is found to be an unconventional in origin. The presence of nodal line $d-$wave superconductivity is manifested from the TF$-\mu$SR and further supported through our theoretical calculation in ThCoC$_{2}$.


A polycrystalline sample of ThCoC$_{2}$ was prepared by arc melting of the stoichiometric amount of Th (3N), Co (4N), and  C (5N) on a water-cooled Cu hearth in a high-purity Ar atmosphere~\cite{Grant}. The ingots were sealed in an evacuated quartz tube and annealed at 1100 $^\circ$C for 336 hr to get better sample quality. X-ray powder diffraction data revealed that ThCoC$_{2}$ is a single phase with space group $Amm2$~\cite{Grant}. The magnetization data were obtained using a Quantum Design Superconducting Quantum Interference Device.  The $\mu$SR experiments were carried out on the MUSR spectrometer at the ISIS Pulsed Neutron and Muon Source of Rutherford Appleton Laboratory, U.K. The $\mu$SR experiments were conducted in ZF-, and TF modes. Spin-polarized muons (100\%) were implanted into the sample, and the positrons from the resulting muon decay were collected in the detector positions either backward or forward of the initial muon-spin direction for the ZF/LF geometry. For TF-measurements the spectrometer was rotated by 90$^\circ$. The powdered sample of ThCoC$_{2}$ was mounted on a silver holder (99.995\%) and the sample holder was placed in a dilution refrigerator which operated in the temperature range 0.1 K~$\le T \le$~3.5 K. Using three sets of orthogonal coils and an active compensation system the stray magnetic fields at the sample position due to the Earth and neighboring instruments are canceled to within 1 $\mu$T. TF-$\mu$SR experiments were performed in the superconducting mixed state in several applied magnetic fields between 10 and 50 mT, due to the low value of $\mu_{0}H_{c2}$. Data were obtained in the field-cooled mode where the magnetic field was applied above $T_c$ and the sample was then cooled down to the lowest temperature. The $\mu$SR  data were analyzed using WiMDA software~\cite{FPW}.   

\begin{figure*}[t]
\centering
\includegraphics[width=\linewidth,height=6.6cm]{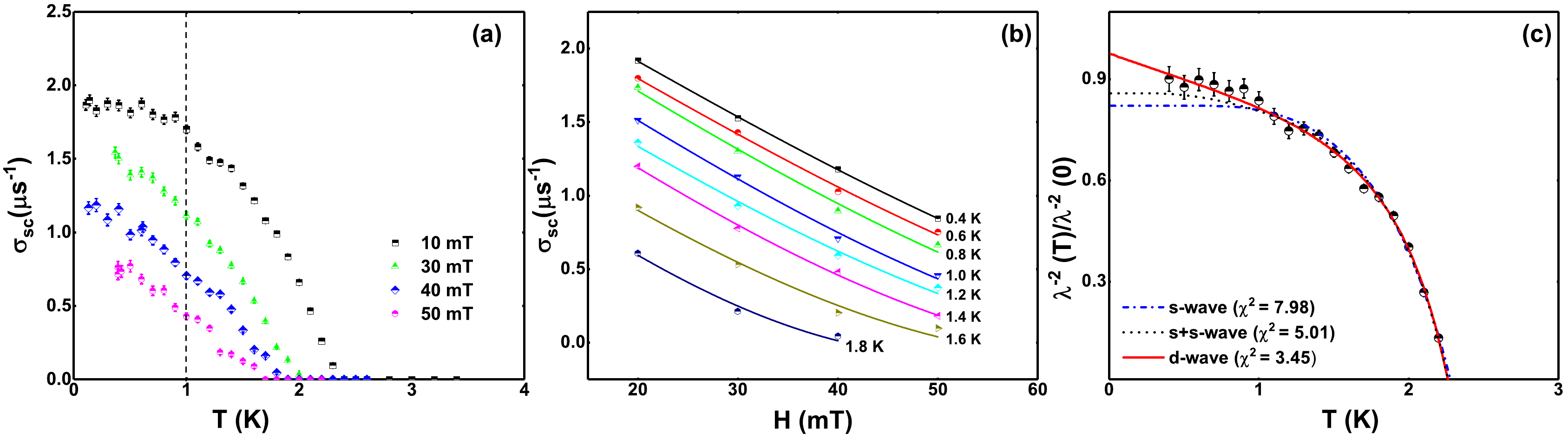}
\caption {(Color online). (a) Temperature dependence of the TF$-\mu$SR spin-depolarization rate collected in a range of fields 10 mT~$\le H \le$~50 mT. The dashed line shows an example of the isothermal cuts used to find the field dependence of $\sigma_{sc}$ at a particular temperature. (b) Field dependence of the muon spin-depolarization rate for a range of different temperatures. The solid lines are the results of fitting the data using Brandt Eq. as discussed in the text. (c)  Temperature dependence of the inverse magnetic penetration depth squared. The lines show the fits using $s-$wave (dashed), $s+s-$wave (dotted) and $d-$wave (solid) gap functions.}
\label{fig2}
\end{figure*}

The superconducting transition temperature $T_c$ = 2.3 K was confirmed by the magnetic susceptibility $\chi(T)$, as shown in Fig.~\ref{fig1}(b). The observed superconducting volume was 100\% of perfect diamagnetism~\cite{Sonier}.  The flux line lattice (FLL) in the vortex state of a type-II superconductor leads to a distinctive field distribution in the sample, which can be detected in the $\mu$SR relaxation rate.  Fig.~\ref{fig1}(c-f) shows the TF-$\mu$SR asymmetry spectra above and below $T_{c}$ in applied field of 10 mT and 30 mT. Below T$_{c}$, the TF-$\mu$SR precession signal decays with time due to the distribution of fields associated with the FLL.  The analysis of TF-$\mu$SR asymmetry spectra was carried out in the time domain using a sinusoidal function damped with Gaussian relaxation plus a nondecaying oscillation that originates from muons stopping in the silver sample holder, $G_{z}(t) = A_{1} \cos (2\pi\nu_{1}t + \phi)\exp (-\frac{\sigma^{2}t^{2}}{2}) + A_{2} \cos(2\pi\nu_{2} t+\phi)$, where $A_1$ and $A_2$ are the sample and background asymmetries, $\nu_{1}$ and $\nu_{2}$ are the frequencies of the muon precession signal from the sample and background signal, respectively with phase angle $\phi$ and $\gamma_\mu/2\pi$ = 135.5 MHz T$^{-1}$ is the muon gyromagnetic ratio.  $\sigma$ is the overall muon depolarization rate; there are contributions from both the vortex lattice ($\sigma_{sc}$) and nuclear dipole moments ($\sigma_{nm}$, which is assumed to be temperature independent) [where $\sigma = \sqrt{(\sigma_{sc}^2+\sigma_{nm}^2}$]. $\sigma_{sc}$ was determined by quadratically deducting the background nuclear dipolar depolarization rate obtained from spectra measured above T$_{c}$, which is shown in Fig.~\ref{fig2}(a).

As can be seen in Fig.~\ref{fig2}(a) that $\sigma_{sc}$ depends on the applied field. Isothermal cuts perpendicular to the temperature axis of the $\sigma_{sc}$ data sets were used to determine the $H$ dependence of the depolarization rate $\sigma_{sc}(H)$ displayed in Fig.~\ref{fig2}(b). Brandt~\cite{Brandt} suggested a useful relation that describe the field dependence for a hexagonal lattice, which is valid over the field range examined in this experiment, $\sigma_{s}[\mu s^{-1}] = 4.83 \times 10^{4}(1-H_{ext}/H_{c2})\times [1+1.2(1-\sqrt{H_{ext}/H_{c2})^3]\lambda^{-2}[nm]}$, to estimate the temperature dependence of the inverse-squared penetration depth, $\lambda^{-2}$ and $\mu_{0}H_{c2}$. The resulting fits to the $\sigma_{sc}(H)$ data are displayed as solid lines in Fig.~\ref{fig2}(b). The temperature dependence of $\lambda^{-2}$ is presented in Fig.~\ref{fig2}(c). We can model the temperature dependence of the $\lambda^{-2}$ by the following equation~\cite{Bhattacharyya1,Adroja1}, $\frac{\lambda^{-2}(T,\Delta_{0,i})}{\lambda^{-2}(0,\Delta_{0,i})}= 1 + \frac{1}{\pi}\int_{0}^{2\pi}\int_{\Delta(T)}^{\infty}(\frac{\delta f}{\delta E}) \times \frac{EdEd\phi}{\sqrt{E^{2}-\Delta(T,\Delta_{i}})^2} \nonumber$, where $f= [1+exp(-E/k_{B}T)]^{-1}$ is the Fermi function. The gap is given by $\Delta_{i}(T,0)=\Delta_{0,i}\delta(T/T_{c})g(\phi)$, whereas g($\phi$) refer to the angular dependence of the superconducting gap function and $\phi$ is the azimuthal angle along the Fermi surface. The $T$ dependence of the superconducting gap is approximated by the relation $\delta(T/T_{c}) = \tanh\{1.82(T_{c}/T-1)]^{0.51}\}$. g($\phi$) is substituted by (a) 1 for an $s-$ or $s+s-$wave gap and (b) $\vert\cos(2\phi)\vert$ for a $d-$wave gap with line nodes.

We have analyzed the temperature dependence of $\lambda^{-2}$ based on three different models, a single-gap isotropic $s-$wave, a line node $d-$wave and two gap $(s + s)-$wave models. The result of the fits using various gap models are shown by lines (dashed, dotted, and solid) in Fig.~\ref{fig2}(c). It is clear from Fig.~\ref{fig2}(c) that the $s-$ or $s+s-$ wave models do not fit the data and gave a larger value of $\chi^2$ = 7.98(3) for $s-$wave and $\chi^2$ = 5.01(1) for $s+s-$wave.  As square root field dependence of the electronic heat capacity suggests a presence of nodes in the superconducting gap symmetry, we therefore fitted our data using a $d-$wave model with line nodes in the gap structure. The $d-$wave model gives the best description of the temperature dependence of $\lambda^{-2}$ of ThCoC$_2$ as the $\chi^2$ value reduced significantly ($\chi^2$ = 3.45(1) for $d$-wave). The $d-$wave fit gives $\Delta(0)$ = 0.77(1) meV and $T_{c}$ = 2.3(1) K. This gives a gap to $T_{c}$ ratio 2$\Delta(0)$/$k_B$$T_{c}$ = 7.8, which is significantly larger than the weak coupling BCS value of 3.53, indicating strong coupling superconductivity in ThCoC$_2$.  Enhanced value of the ratio is also observed in other NCS superconductors such as Re$_6$Zr~\cite{Singh} with 2$\Delta(0)$/$k_B$$T_{c}$ = 4.2, BiPd~\cite{Sun1} with 2$\Delta(0)$/$k_B$$T_{c}$ = 3.8,  La$_7$Ir$_3$~\cite{Barker} with 2$\Delta(0)$/$k_B$$T_{c}$ = 3.81, K$_2$Cr$_3$As$_3$~\cite{Adroja2015} with 2$\Delta(0)$/$k_B$$T_{c}$ = 6.4 and  Cs$_2$Cr$_3$As$_3$~\cite{Adroja2017} with 2$\Delta(0)$/$k_B$$T_{c}$ = 6.0. Our ZF$-\mu$SR study on ThCoC$_2$ shows similar relaxation rate above and below $T_c$, indicting that time reversal symmetry is preserved, see details in Ref.~\cite{amitavaTh}.

\begin{figure*}[t]
\centering
\includegraphics[width=\linewidth]{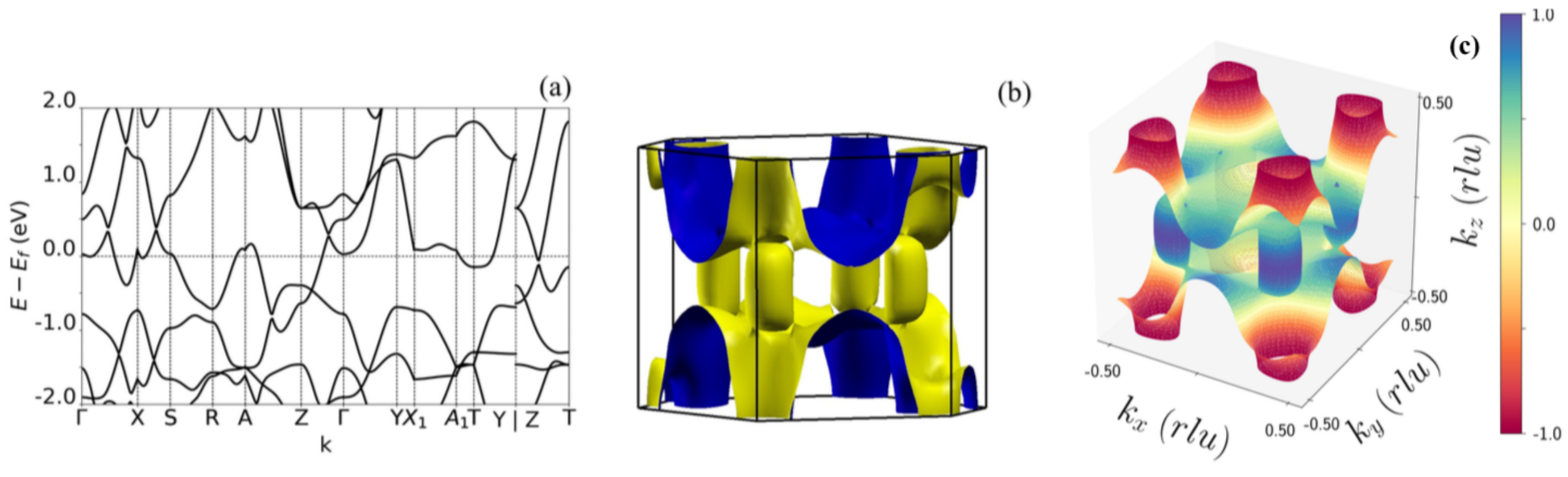} 
\caption{(a) Computed result of DFT band structure is plotted along the standard high-symmetric directions for this crystal structure. (b) Corresponding Fermi surface topology. The color in this plot does not have any specific meaning. (c) Computed pairing eigenfunction plotted in a color map of red (negative) to yellow (nodes) to blue (positive) color. The result is overlaid on the corresponding FS. We immediately observe the presence of nodal lines on the $k_z=\pm\pi/2$ planes (visualized by yellow contour).}
\label{fig3}
\end{figure*}


Now we discuss our theoretical calculations. For the first-principles electronic structure calculation, we used the Vienna Ab-initio Simulation Package (VASP)~\cite{VASP} and use the Perdew-Burke-Ernzerhof (PBE) form for the exchange-correlation functional~\cite{PBE}. To deal with the strong correlation effect of the $d$-electrons of the Th and Co atoms, we employed LDA+$U$ method with $U = 5 eV$ on both atoms. We have recalculated the band structure with spin-orbit coupling, and no considerable change is obtained in the low-energy bands of present interests. The computed band structure is shown in Fig.~\ref{fig3}(a). We notice that only a single band osculates around the Fermi level. This band is dominated by Co-$d$-orbital with hybridization with the C-$p$ orbitals (with about 10\% contributions from the C-$p$ states to the low-energy band structure). The corresponding Fermi surface (FS) is shown in Fig.~\ref{fig3}(b). The three-dimensional FS topology is quite interesting in this system, exhibiting a hole-like pocket around the zone corner (`A' point), but an electron-like topology around the `T'-point. In addition, as we scan across the $k_z$-direction, we notice that the Fermi pocket near the $k_z=\pm\pi$-plane first closes at discrete Fermi-points, and then reopens into another pocket around some non-high-symmetric point. Such an interesting FS topological change in the 3D Brillouin zone (BZ) within a single band model implies the presence of saddle-points in the electronic structure around the Fermi level. Hence one obtains a large density of states near the Fermi level, which is desirable for superconductivity. In addition, due to multiply-pockets features in the FS, one expects large FS nestings, and hence unconventional superconductivity.


We perform a pairing symmetry calculation using single-band Hubbard model. By carrying out a perturbative expansion of the Hubbard interaction, we obtain an effective pairing potential in the single and triplet pairing channels~\cite{spinfluc,pairpot}
\bea
\label{pairpot}
V_{\uparrow\downarrow}(\mathbf{q})&=&\frac{U}{2}\left[1+3U\chi^{s}(\mathbf{q})\right]+\frac{U}{2}\left[1-U\chi^{c}(\mathbf{q})\right],\nonumber\\
V_{\uparrow\uparrow}(\mathbf{q})&=&\frac{U}{2}\left[1-U\chi^{s}(\mathbf{q})\right]+\frac{U}{2}\left[1-U\chi^{c}(\mathbf{q})\right].\\
\eea
Here $\chi^{s/c}(\mathbf{q})$ are the spin and charge susceptibilities, decoupled within the random-phase approximation (RPA), as $\chi^{s/c}({\bf q}) = \chi^{0}({\bf q})/(1\mp U\chi^{0}({\bf q}))$ with $\chi^{0} ({\bf q})$ being bare static Lindhard susceptibility. For such a potential, the BCS gap equation (for details see Supplementary materials) becomes
\be
\Delta_{\mathbf{k}}=-\lambda \sum_{\mathbf{k^{'}}}V(\mathbf{k}-\mathbf{k^{'}})\Delta_{\mathbf{k^{'}}}.
\label{Paireig}
\ee
This is an eigenvalue equation with the pairing strength $\lambda$ denoting the eigenvalue. For $V>0$, a positive eigenvalue $\lambda$ can commence with the corresponding eigenfunction $\Delta_{\bf k}$ which changes sign as ${\rm sgn}[\Delta_{\bf k}]=- {\rm sgn}[\Delta_{\bf k'}]$ promoted by strong peak(s) in $V$ at ${\bf Q}=\mathbf{k}-\mathbf{k^{'}}$. Looking into the origin of $V$ in Eq.~\eqref{pairpot}, we notice that $V$ inherits strong peaks from that in $\chi^{s/c}$, which is directly linked to the FS nesting feature embedded in $\chi^0$. In this way, the pairing symmetry of a system is intricately linked to the FS nesting instability. This theory of spin-fluctuation driven superconductivity consistently links between the observed pairing symmetry and FS topology in many different unconventional superconductors~\cite{spinfluc,pnictides,HF}.

For the present FS topology presented in Fig.~\ref{fig3}(c), we find that the nesting is dominant between the two FS pockets centered at the $k_z=0$, and $k_z=\pm \pi$ planes. This nesting is ingrained within the above pairing potential $V$, and naturally dictates a dominant pairing channel that changes sign between these two $k_z$-planes. In fact, our exact evaluation of the pairing eigenvalue gives the dominant pairing channel to be a spin-singlet pairing with $d_z$-wave pairing channel $\Delta_{\bf k}=\cos(k_z)$ with an eigenvalue of $\lambda\sim 0.2$. The computed pairing eigenfunction is shown in Fig.~\ref{fig3} on the FS in a color plot. The computed pairing channel does not acquire any in-plane anisotropy, and hence a nodal line on the $k_z=\pm \pi/2$. The experimental data fitted with the same nodal-like pairing channel indeed gives a better fit compared to other pairing function. 


In summary, we have investigated the gap symmetry of ThCoC$_{2}$ using transverse field $\mu$SR. Our $\mu$SR analysis confirmed that a nodal line $d-$wave model fits better than a single or two gaps isotropic $s-$wave model to the observed temperature dependence of the penetration depth. This finding is in agreement with the field dependent heat capacity, $\gamma(H) \sim \sqrt{H}$.  The finding of the nodal line gap is further supported through our theoretical calculations. This work paves the way for further studies of the large numbers of unexplored NCS compounds with the CeNiC$_2$ type structure for the hunt of the unconventional superconductor.


We would like to thanks, Prof. A. M. Strydom for interesting discussions. AB would like to acknowledge DST India, for Inspire Faculty Research Grant (DST/INSPIRE/04/2015/000169), FRC of UJ, NRF of South Africa and ISIS-STFC for funding support. DTA and ADH would like to thank CMPC-STFC, grant number CMPC-09108. KP would like to acknowledge DST India, for Inspire Fellowship (IF170620). TD acknowledges financial support from the Science and Engineering Research Board (SERB) of the Department of Science \& Technology (DST), Govt. of India for the Start-Up Research Grant (Young Scientist), and also benefited from the financial support from the Infosys Science foundation under Young Investigator Award.

\newpage
\section{Supplementary Material}
\subsection{Evidence of Nodal Line in the Superconducting Gap Symmetry of Noncentrosymmetric ThCoC$_{2}$} 
\section{I. Time reversal Symmetry}

The time evolution of the ZF$-\mu$SR of ThCoC$_2$ is shown in Fig.S1 for $T$ = 0.4 K and 3.5 K. In these relaxation experiments, any muons stopped on the silver sample holder gave a time independent background. No signature of precession is visible (either at 0.4 K or 3.5 K), ruling out the presence of a sufficiently large internal magnetic field as seen in magnetically ordered compounds. The only possibility is that the muon$-$spin relaxation is due to static, randomly oriented local fields associated with the electronic and nuclear moments at the muon site. The ZF$-\mu$SR data are well described in terms of the damped Gaussian Kubo-Toyabe (KT) function, $G_{z2}(t) =A_1 G_{KT}(t)e^{-\lambda t}+A_{bg}$ where $G_{KT}(t) = \left[\frac{1}{3}+\frac{2}{3}(1-\sigma_{KT}^2t^2)e^{{\frac{-\sigma_{KT}^2t^2}{2}}}\right]$ is Gaussian Kubo-Toyabe function, $\lambda$ is the electronic relaxation rate, $A_1$ is the initial asymmetry, $A_{bg}$ is the background.  The parameters $\sigma_{KT}$, $A_1$, and $A_{bg}$ are found to be temperature independent.  It is evident from the ZF$-\mu$SR spectra that there is no noticeable change in the relaxation rates at 3.5 K ($\ge T_{\bf c}$) and 0.4 K ($\le T_{\bf c}$). This indicates that the time-reversal symmetry is preserved upon entering the superconducting state. $\sigma_{KT}$ accounts for the Gaussian distribution of static fields from nuclear moments (the local field distribution width $H_\mu$ = $\sigma/\gamma_\mu$, with muon gyromagnetic ratio $\gamma_\mu$ = 135.53 MHz/T). The fits of $\mu$SR spectra in Fig.S1 by the decay function gave $\sigma_{KT}$ = 0.31(7) $\mu s^{-1}$ and $\lambda$ = 0.12(6) $\mu s^{-1}$ at 3.5 K and $\sigma_{KT}$ = 0.30(7) $\mu s^{-1}$ and $\lambda$= 0.11(9) $\mu s^{-1}$ at 0.4 K. The fits are shown by the solid lines in Fig.S1. Since within the error bars both $\sigma_{KT}$ and $\lambda$ at $T < T_{\bf c}$ and $T > T_{\bf c}$ are similar, there is no evidence of time-reversal symmetry breaking in ThCoC$_2$. This result also confirm the absence of any kind of magnetic impurity in our sample.

\begin{figure*}[t]
\vskip -0.5 cm
\centering
\includegraphics[width=0.6\linewidth]{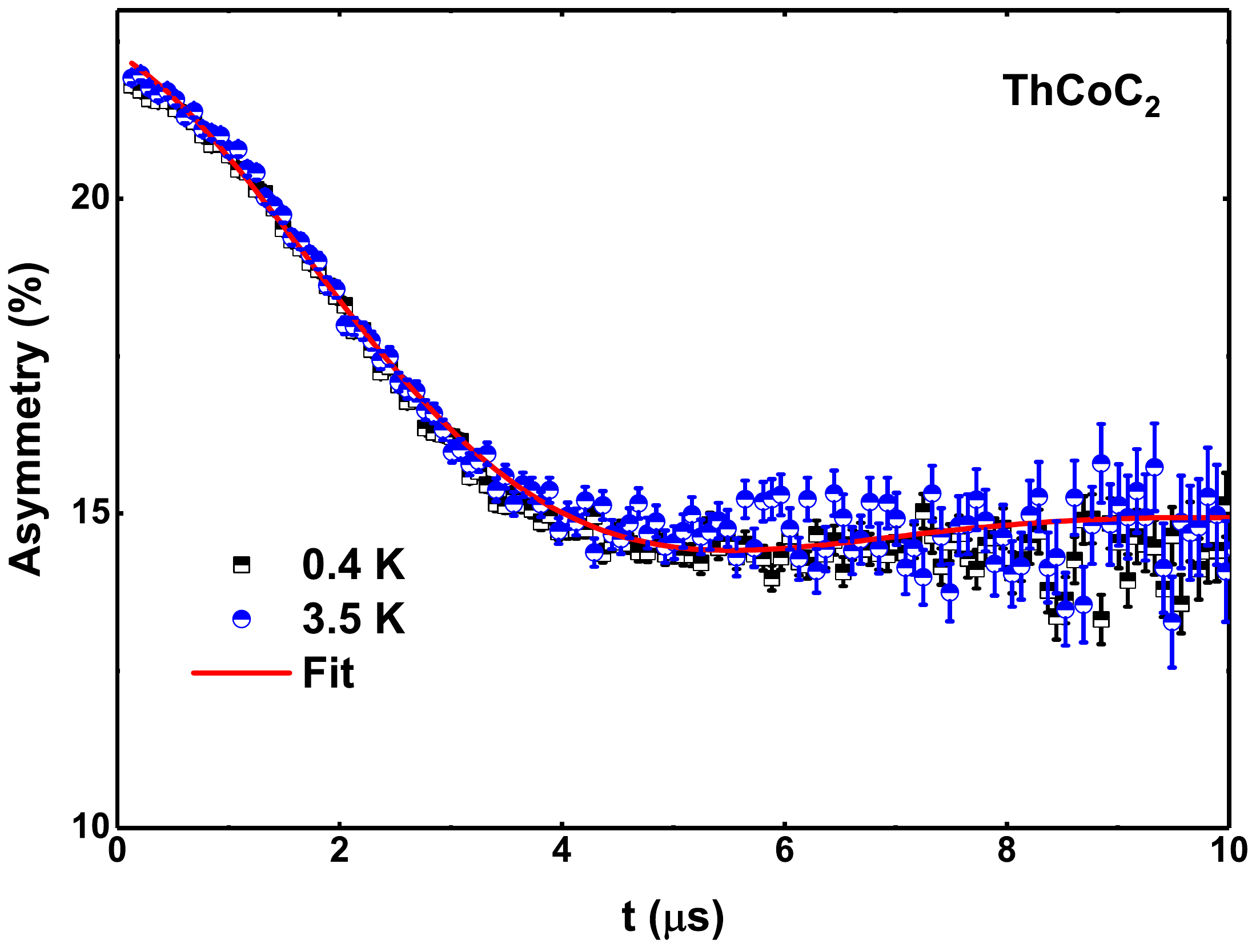}
\caption {(Color online) (a)  Time dependence asymmetry spectra at 0.4 K and 3.5 K measured in zero magnetic field of ThCoC$_2$ with $T_{\bf c}$ = 2.3 K. The lines are the results of the fits. The absence of extra relaxation below $T_{ c}$ indicates no internal magnetic fields and, consequently, suggests that the superconducting state preserves time reversal symmetry.}
\label{fig4}
\end{figure*}

\section{II. DFT calculation}
For the first-principles electronic structure calculation, we used the Vienna Ab-initio Simulation Package (VASP)~\cite{VASP} and use the Perdew-Burke-Ernzerhof (PBE) form for the exchange-correlation functional~\cite{PBE}. The projector augmented wave (PAW) pseudo-potentials are used to describe the core electrons~\cite{PAW}. Electronic wave-functions are expanded using plane waves up to a cut-off energy of 500 eV. The Monkhorst-Pack k-mesh is set to $14\times14\times14$ in the Brillouin zone for the self-consistent calculation. All atoms are relaxed in each optimization cycle until atomic forces on each atom are smaller than 0.01 $eV \AA$. The lattice constants are obtained by relaxing the structure and with total energy minimization. Our obtained relaxed lattice parameters are $a$ = 3.8493${\AA}$, $b$ = 3.8493$\AA$, $c$ = 3.9462$\AA$, which are close to the experimental values~\cite{ExpLat1}. To deal with the strong correlation effect of the $d$-electrons of the Th and Co atoms, we employed LDA+$U$ method with $U = 5 eV$ on both atoms. We have recalculated the band structure with spin-orbit coupling, and no considerable change is obtained in the low-energy bands of present interests.

\section{II. Details of pairing eigenvalue calculations}

Our calculation of pairing interaction and pairing eigenvalue originating from spin-fluctuation is done by directly including the DFT band structure in a three-dimensional Brillouin zone (BZ). We start with a single-band Hubbard model, as dictated by the DFT calculation, which is given by
\bea
H=\sum_{\mathbf{k},\sigma=\uparrow,\downarrow}\xi_{\mathbf{k}}c^{\dagger}_{\mathbf{k}\sigma}c_{\mathbf{k}\sigma}+U\sum_{\mathbf{k},\mathbf{k}^{'},\mathbf{q}}c^{\dagger}_{\mathbf{k}\uparrow}c_{\mathbf{k}+\mathbf{q}\uparrow}c^{\dagger}_{\mathbf{k}^{'}\downarrow}c_{\mathbf{k}^{'}-\mathbf{q}\downarrow},\nonumber\\
\eea
where $c^{\dagger}_{\mathbf{k}\sigma} (c_{\mathbf{k}\sigma})$ is the creation (annihilation) operator for non-interacting electron with momentum ${\bf k}$, and spin $\sigma=\uparrow/\downarrow$, and $\xi_{\bf k}$ is the corresponding band structure, taken directly from the DFT calculation. The second term is the Hubbard interaction, written in the band basis with the onsite potential $U$. Perturbative expansion of the Hubbard term in the spin-singlet and spin-triplet pairing channels yields,
\bea
H&=&\sum_{\mathbf{k},\sigma=\uparrow,\downarrow}\xi_{\mathbf{k}}c^{\dagger}_{\mathbf{k}\sigma}c_{\mathbf{k}\sigma}\nonumber\\
&&+\sum_{\mathbf{k}\mathbf{k}^{'},\sigma\sigma'}V_{\sigma\sigma'}(\mathbf{k}-\mathbf{k^{'}})c^{\dagger}_{\mathbf{k}\sigma}c^{\dagger}_{-\mathbf{k}\sigma^{'}}c_{-\mathbf{k^{'}}\sigma^{'}}c_{\mathbf{k}'\sigma}.
\label{pairH}
\eea
Here $V_{\sigma\sigma'}$ refers to the spin singlet and triplet case when $\sigma=\mp\sigma$, respectively. The corresponding (static) pairing potentials are given in Eqs.(2-3) in the main text. The bare susceptibility is 
\be
\chi^0(\mathbf{q},\omega)=-\sum_{\mathbf{k}}\frac{f(\xi_{\mathbf{k}})-f(\xi_{\mathbf{k}+\mathbf{q}})}{\omega-\xi_{\mathbf{k}}+\xi_{\mathbf{k}+\mathbf{q}}+\iota \delta},
\ee
where $f$ is the corresponding Fermi-function, and $\delta$ is an infinitesimal number. Since we are interested in the static limit of the pairing potential, the imaginary part of the above particle-hole correlator does not contribute and we obtain a real pairing potential $V$. In the RPA channel, the spin and charge channels are decoupled, and due to different denominators $1\mp U\chi^0$, the spin channel is strongly enhanced while simultaneously charge channel is suppressed. The method is a weak to intermediate coupling approach and thus the value of $U$ is restricted by the non-interacting bandwidth. We use $U$ = 400 meV in our numerical calculation and the result is reproduced with different values of $U$. We note that this value of $U$ is defined in the band basis, while the LDA $U$ is for the orbitals. The $k$-dependent of the pairing eigenfunction is dictated by the anisotropy in $V({\bf q})$ which is directly related to the bare FS nesting character, this result remains unaffected by a particular choice of $U$ in the weak to intermediate coupling range. With increasing $U$, the overall strength of $V$ mainly increases and thus the pairing eigenvalue $\lambda$ also increases. Since in this work, we are primarily interested in uncovering the pairing eigenfunction, the value of $U$ remains irrelevant in this coupling strength. 

From interacting Hamiltonian, we define the SC gap equation as
\begin{align}
\Delta_{\mathbf{k}}&=-\sum_{\mathbf{k^{'}}}V_{\sigma\sigma'}(\mathbf{k}-\mathbf{k^{'}})\langle c_{\mathbf{-k}'\sigma}c_{\mathbf{k}'\sigma'}\rangle,\\
&= -\sum_{\mathbf{k^{'}}}V_{\sigma\sigma'}(\mathbf{k}-\mathbf{k^{'}})\frac{\Delta_{\mathbf{k^{'}}}}{2\xi_{\mathbf{k^{'}}}} \tanh{\Big(\frac{\xi_{\mathbf{k^{'}}}}{k_BT_c}\Big)}
\end{align}
In the limit $T\to 0$ we have $\langle c_{-{\bf k}\sigma}c_{\mathbf{k}\sigma'}\rangle\to \lambda \Delta_{\mathbf{k^{}}}$, leading to an eigenvalue equation as
\begin{align*}
\Delta_{\mathbf{k}}=-\lambda \sum_{\mathbf{k^{'}}}V_{\sigma\sigma'}(\mathbf{k}-\mathbf{k^{'}})\Delta_{\mathbf{k^{'}}}.
\end{align*}
We solve this equation separately for spin singlet ($\sigma=-\sigma'$), and spin triplet ($\sigma=\sigma'$) cases. Since we restrict ourselves to the static limit, the above equation is solved for only Fermi momenta $\mathbf{k}_F$, $\mathbf{k}'_F$ in which $V_{\sigma\sigma'}(\mathbf{k}_F-\mathbf{k}'_F)$ becomes a $n\times n$ matrix with $n$ being the number of discrete Fermi momenta considered in a 3D BZ.

\end{document}